\documentstyle{article}
\begin{document}

\title{Hartree-Fock Approximation for the $Ab~Initio$ No-Core Shell Model}

\author{Mahmoud A. Hasan$^{1}$, James P. Vary$^2$ and Petr 
Navr\'{a}til$^3$\\
(1) Department of Physics, Applied Science University\\
Amman, Jordan\\
\noindent (2) Department of Physics and Astronomy, Iowa State University\\
Ames, Iowa 50011\\
\noindent (3) Lawrence Livermore National laboratory, Livermore, California 
94550  \\
}
\date{\today}
\maketitle

\begin{abstract}
The spherical Hartree-Fock approximation is applied to
the $ab~ initio$ no-core shell model, with a realistic effective
nucleon-nucleon interaction in order to investigate the range of its utility. 
Hartree-Fock results for binding energies, one-body density distributions 
and occupation probabilities are compared with results from exact 
diagonalization in similar model spaces. We
show that this mean field approximation, especially with second order 
corrections, is able to provide some useful approximatons for
$^4He$ and $^{16}O$.
We also explore the physical insights provided by the Hartree-Fock 
results for single-particle properties such as spin-orbit splittings.  We find
single particle state ordering consistent with the 
phenomenological shell model.

\end{abstract}

\section{Introduction}

Recently, the $ab~ initio$ no-core shell model (NCSM) has been applied 
with realistic effective nucleon-nucleon (NN) interactions  to light nuclei 
up to A=12 \cite {Zheng95, Petr96, PRL00, PRC00}.  With the need to extend 
to heavier systems and to incorporate improvements such as effective 
and real three-body forces in ever larger model spaces, the prospect for 
near-term results is limited by present day computational resources.  
In light of this situation, there is a need for approximate methods to 
extend the $ab~ initio$ NCSM to heavier systems with a wide range of 
observables to compare with experiment.

Hartree-Fock is a proven tool for semi-realistic interactions 
for even the heaviest of nuclei \cite {pring80} and is sufficiently flexible to handle many-body
forces through the role of delta excitations \cite {c5, c6, c7, c8}.
It is also the starting point for practical many-body methods used extensively in 
heavier systems \cite{pring80}.
One of the new questions to address is the quantitative accuracy 
of Hartree-Fock itself when using the
latest theories for effective interactions based on realistic NN potentials.  
Here, we provide an initial comparison in light nuclei which leads us to 
conclude that care must be exercised in
the use of the mean field approach with these newer effective Hamiltonians. 
The size of second order corrections is found to be a useful gauge of the 
utility of the mean field method in the present comparison.

Two recent efforts \cite {SSR00,Coraggio03}, taken together, 
show that  the higher order corrections to Hartree-Fock are rather sensitive to the 
choice of Hamiltonian.  On the one hand, using phenomenological interactions 
Ref. \cite {SSR00} presents
higher order corrections that are significantly smaller than those we obtain.  
These phenomenological interactions  provide a good description of many experimental 
observables within the mean field approach. It is not clear whether these phenomenological 
interactions would provide good descriptions of experiment in the NCSM approach or 
any other $ab~ initio$ method.  On the other hand, using a new method \cite {Bogner} to 
develop a realistic low-momentum nucleon-nucleon potential, called "$V_{low-k}$", 
Ref. \cite {Coraggio03}  evaluated the Hartree-Fock results for $^{16}O$ and $^{40}Ca$
including corrections through third order.  For $^{16}O$ the second order corrections 
of Ref. \cite {Coraggio03} are somewhat larger than those we obtain.  On the other
hand their Hartree-Fock results through third order are in better agreement with 
experiment.  We discuss further the differences between our results and those of Ref. 
\cite {Coraggio03} in Sec. 4.

Of course, there is a long history going back to Brueckner, of merging the mean-field
method with non-relativistic effective potentials (G-matrix) derived 
from NN interactions \cite {Day67, Negele70, Davies71}. 
The conclusion of this extensive  set of research is that such Hamiltonians 
underbind nuclei by about 1-3 MeV  per nucleon.  The tendency of the results 
is to have a root-mean-square radius ($r_{rms}$) which is too small  compared to experiment 
whenever the binding energy approaches the  experimental value ("Coester line").

Until recently, these extensive results left open the possibility that the mean-field
method along with selected higher-order corrections, included by various means, was not a
sufficiently accurate approach.  However, with the advent of very precise methods to solve the
many-fermion problem  for light nuclei, there appears to be a good consensus now that the 
deficiency lies with the Hamiltonian itself.  That is, we need true many-body forces  
to resolve the discrepancies between theoretical and experimental ground state ($GS$) properties.

Thus, we can easily imagine that properly constructed Hamiltonians, consisting of 
bare NN and NNN interactions, renormalized for large but finite basis spaces, could provide high precision
descriptions of a wide variety of low-energy properties of nuclei.  We then require many body techniques
that propel the applications in all nuclei, not just light nuclei.  Given the recent advances in constructing
such effective Hamiltonians, we may begin to re-assess the utility of mean field methods and their extensions
for these purposes.
 
Our intent here is rather focused on a particular set of issues. We aim to examine the utility of the
mean-field method with one of the more recent effective interaction approaches.  We need to do this if 
we are to open the  door to incorporating delta excitations as one of the important mechanisms for many
body  forces in nuclei and if we are to proceed to heavier nuclei retaining predictive power.  
Indeed, we have been  working in this direction for some time
\cite {c5, c6, c7, c8} with effective Hamiltonians based  on G-matrices augmented by N-$\Delta$ and
$\Delta-\Delta$ interactions.

In the present effort, we have two specific goals: first, to compare spherical Hartree-Fock 
(SHF) with the $ab~initio$ NCSM in light nuclei where both methods are solvable with newly developed
effective Hamiltonians in order to determine the quantitative accuracy of SHF and the associated conditions; 
second, to extract additional physical insights from SHF with these realistic effective Hamiltonians as a
complement  to the NCSM results.

\section{The Effective Hamiltonian}

The $ab~ initio$ approach in shell-model studies 
of the nuclear many-body problem starts 
\cite {Zheng95, Petr96, PRL00, PRC00} with the 
intrinsic two-body Hamiltonian for the $A$-nucleon system, i.e., 

\begin{equation}
H=  \sum_{i<j}^{A} \left [ T_{ij} + V_{ij} \right ]
\end{equation}

\noindent with $T_{ij}$ the relative kinetic energy
between NN pairs and $V_{ij}$ the $NN$ 
interaction including the Coulomb interaction between protons. 
We ignore three-body interactions in the present effort.
For the purposes of evaluating an
effective Hamiltonian we modify it by adding (and later subtracting)
the center-of-mass harmonic-oscillator (HO) Hamiltonian, 

\begin{equation}
H^{\Omega}_{cm} = \frac{\vec{P}^2}{2Am} + \frac{1}{2}Am \Omega^2 \vec{R}^2 
\end{equation}

\noindent with $m$ the nucleon mass, $\vec{P}=\sum_{i=1}^{A} \vec{p_i}$,
and $\vec{R}=(1/A)\sum_{i=1}^{A} \vec{r_i}$. 
 
This addition/subtration of a single particle potential, 
first introduced by Lipkin \cite{Lipkin}, 
helps our overall convergence when working in a 
HO set of basis states. We emphasize that it is 
important to ensure, as we do, that the $intrinsic$ properties of 
the many-body 
system are not affected by the center-of-mass term. The 
modified Hamiltonian, thus, acquires a  dependence on the 
HO frequency $\Omega$, and can then be written as

\begin{equation}
H_A^{\Omega}= \sum_{i=1}^{A} \left [ \frac{\vec{p_i}^2}{2m} + \frac{1}{2}m 
\Omega^2 \vec{r_i}^2 \right ] 
+ \sum_{i<j}^{A} \left [ V_{ij} - \frac{m \Omega^2}{2A}(\vec{r_i} - 
\vec{r_j})^2 \right ]
\end{equation}

\noindent Our shell-model calculations are performed in a 
model space defined by a projection operator $P$, with the complementary space 
(i.e. the excluded space) 
defined by the projection operator $Q=1-P$. Furthermore, due to its strong 
short-range part, the realistic 
nuclear interaction in Eqs. (1) and (3) will yield pathological results unless 
we derive a model-space dependent $effective$ Hamiltonian: 

\begin{equation}
H^{\Omega}_{eff}= \sum_{i=1}^{A} P \left [ \frac{\vec{p_i}^2}{2m} + 
\frac{1}{2}m \Omega^2 \vec{r_i}^2 \right ] P 
+  P \left [ V_{eff}  \right ] P
\end{equation}

\noindent The effective interaction appearing in Eq. (4) is, in general, 
an $A$-body interaction, and, 
when it is obtained without any approximations, the model-space Hamiltonian 
provides an identical 
description of a subset of states as the exact original 
Hamiltonian \cite{LS1,LS2}. 

From among the
eigenstates  of the Hamiltonian (4), it is necessary to choose 
only those that correspond to the same
center-of-mass energy.  This can be achieved by working in a 
complete $N_{max} \hbar \Omega$ model space, and
then by shifting the  center-of-mass eigenstates with energies 
greater than $\frac{3}{2}\hbar \Omega$
(representing spurious  center-of-mass motion) upwards in the energy 
spectrum. We do this by adding 
$(\beta -1)PH^{\Omega}_{cm}P$ to and subtracting $\beta \frac{3}{2} \hbar 
\Omega P$ from equation (4) above. One unit of $H_{cm}$ has 
already been acquired, as mentioned above \cite{PRC00}. The resulting shell-model 
Hamiltonian takes the form

\begin{eqnarray}
H^{\Omega}_{eff~\beta} &=& \sum_{i<j}^{A} P \left [ \frac{(\vec{p_i}-
\vec{p_j})^2}{2Am} + \frac{m \Omega^2}{2A} 
(\vec{r_i}-\vec{r_j})^2 \right ] P\\  
&& +  P \left [ V_{eff}  \right ] P
+ \beta P(H^{\Omega}_{cm}-\frac{3}{2}\hbar \Omega)P\nonumber
\end{eqnarray}

\noindent where $\beta$ is a sufficiently large positive parameter. 
When applied in a complete 
$N_{max} \hbar \Omega$ model space, this procedure removes the 
spurious center-of-mass motion exactly, and has no 
effect on the intrinsic spectrum of states with the lowest center-of-mass 
configuration \cite{Petr96}.

In principle, the effective interaction introduced in Eqs. (4) and (5) above 
should reproduce exactly 
the full-space results in the model space for some subset of states. 
Furthermore, an $A-$body effective 
interaction is required for an $A-$nucleon system. In practice, however, the 
effective interaction cannot be calculated exactly, and it is approximated 
with a two-body effective interaction determined for a 
two-nucleon subsystem of the A-nucleon system.   More recently, it has been possible
to extend the effective interaction to the three-body cluster level \cite{nav02}.

In this work, we follow the procedure described in Refs. \cite{Petr96, 
PRL00, PRC00}
in order to construct the  two-body effective interaction. The procedure
employs the Lee-Suzuki
\cite{LS1} similarity transformation method, which yields an 
interaction in the form 

\begin{equation}
P_2V_{eff}P_2=P_2VP_2 + P_2VQ_2\omega P_2,\
\end{equation}

\noindent with $\omega$ the transformation operator satisfying 
$\omega=Q_2\omega 
P_2$, and $P_2$, 
$Q_2=1-P_2$ operators that project on the two-nucleon model and complementary 
spaces, respectively. 
Note that we distinguish between the two-nucleon system projection operators 
$P_2,~Q_2$ and the 
$A-$nucleon system projection operators $P,~Q$.  
The choice of $P_2$ is fixed by the choice of $P$.  The remaining detailed steps
to obtain the non-Hermitian form of $H_{2eff}$ follow earlier work 
\cite {PRL00, PRC00,LS1,LS2}.

The final Hermitian form,
$\bar{H}_{2eff}$, is obtained by applying a similarity
transformation determined from  the metric operator 
$P_2(1+\omega^{\dagger}\omega)P_2$ \cite{LS2}:

\begin{equation}
\bar{H}_{2eff}=[P_2(1+\omega^{\dagger}\omega)P_2]^{1/2}H_{2eff}
[P_2(1+\omega^{\dagger}\omega)P_2]^{-1/2} \
\end{equation}

The two-body effective interaction used in the present 
calculations is determined from this two-nucleon 
effective Hamiltonian as $V_{2eff}=\bar{H}_{2eff}-H^{\Omega}_{02}$
where $H^{\Omega}_{02}$ is the relative oscillator Hamiltonian for 
two particles. The 
resulting two-body effective interaction 
$V_{2eff}$ depends on $A$, on the HO 
frequency $\Omega$, and on $N_{max}$, the maximum many-body 
HO excitation energy (above the lowest 
configuration) defining 
the $P-$space. Furthermore, as discussed earlier, when 
used in the shell-model Hamiltonian (5), it results in the 
factorization of our 
many-body wavefunction into 
a product of a center-of-mass $\frac{3}{2}\hbar \Omega$ component times an 
intrinsic component, which allows 
exact correction of any observable for spurious center-of-mass effects, thus 
preserving translational invariance. 
This feature distinguishes our approach from most phenomenological shell-model 
studies that involve multiple HO shells.

So far, the most important approximation used in our approach is the 
neglect of contributions coming from higher 
than two-body clusters to our effective Hamiltonian. 
In the NCSM 
the inclusion of  a three-body effective interaction 
has been accomplished for $0s$- and $0p$- shell nuclei 
\cite{nav02,Petr98,Petr00} though computational needs
increase rapidly.   For SHF it is straightforward, in principle, to carry out
investigations with multi-body effective Hamiltonians.

While the preservation of translational invariance in the NCSM is exact, 
this is not the case in our SHF approach.  It is well known that projection 
before variation is desirable for obtaining optimized solutions respecting a  
given symmetry not already guaranteed by mean field basis selection.  
Thus,  it is possible to implement an exact treatment of translational 
invariance within Hartree Fock \cite{schmid}.   
Here, instead, we set $\beta=0$ in Eq. (5)
and solve the conventional SHF problem.  Since we introduce 
a SHF model space truncation and we solve for a single
slater determinant, our SHF results acquire center of mass 
(c.m.) motion dependence. 
Thus, our SHF rms radius and one-body density will have c.m. 
wavefunction smearing and we approximately correct for this in our 
rms radius results below.

The preservation of translational invariance in the effective Hamiltonian
brings about a very interesting set of consequences for the 
mean field single particle energies \cite {Jaqua92}.   Hence some care
must be exercised in their interpretation and in comparison with results
from other Hamiltonians.  

For cases with either a purely intrinsic Hamiltonian (no 1-body component)
or a Hamiltonian with a 1-body component plus intrinsic terms, we can list the common
features. First, the single-particle energy
is the eigenvalue of a mean-field one-body Hamiltonian equation derived
from the application of the variational principle to the initial Hamiltonian.  
Second, it is this 1-body self-consistent field problem that defines the leading 
order mean field single particle properties with which  higher order corrections
 are to be evaluated.  Thus, in either case, it is the resulting single particle energies 
that appear in the energy denominators of higher order perturbation theory.  
Also, the associated single particle wavefunctions are used to evaluate the 
matrix elements of the perturbative corrections.  Third, in neither case may 
these single particle energies be directly compared with experiment without 
considering the role of rearrangement.  The single particle energies and their  
associated rearrangement effects are not independent of each other and
neither corresponds directly to an observable. Hence they may differ 
significantly between the two types of Hamiltonians.

We will now see why our single particle energies differ substantially 
from those obtained with a combination of a one-body
and a two-body Hamiltonian as conventionally employed.  We will also see that 
we can easily extract results for spin-orbit splittings that are not markedly 
different from results of other approaches.

In particular, for our purely intrinsic effective Hamiltonians, 
we obtain a simple relationship between the
Hartree Fock energy and the single particle energy:

\begin{eqnarray}
E_{HF} = 0.5 \sum \epsilon_A (2j_A+1)
\end{eqnarray}
where we signify orbits occupied in the slater determinant by capital Roman letters.
One can easily verify this relationship with the SHF results presented below.   This 
relationship was already evident from thermal mean field studies using 
pure 2-body no-core Hamiltonians \cite {Bozzolo88} and was examined 
in some detail in Ref \cite {Jaqua92}.

Herein lies an important bridge for comparing with experimental 
single particle energies.  In particular, by neglecting rearrangement effects, as 
is traditional when comparing mean field results with experimental states in 
neighboring odd-mass nuclei, we see that the HF excitation energy is only 
one-half of the difference in the single particle energies of a promoted
particle.  Thus if the difference in the single particle energies between a particular
single particle state above the Fermi suface and one below, 
obtained with our intrinsic Hamiltonian, is 20 MeV for example, then the  Hartree Fock 
energy of the associated Hartree Fock excited state is just 10 MeV 
above the Hartree Fock ground state.

To make the comparison with experiment more precise, we would need to
carry out an evaluation of the rearrangement energy which takes us beyond 
the scope of the present effort.  However, insofar as rearrangement effects may 
be neglected with both classes of Hamiltonians discussed above, it seems 
reasonable to compare one-half of our single-particle energy differences 
with full single particle energy differences obtained from Hamiltonians 
having a 1-body component.  On this basis, we will compare quantitatively in Sec. 4 
the example of spin-orbit splittings obtained with our 
Hamiltonian and the results of Ref. \cite {Coraggio03} which employed a 
Hamiltonian with a 1-body component.

As all Hartree Fock energies are then proportional to the single particle energies, we
can obtain evidence on shell properties (single particle state ordering and relative 
spacings) from the mean field results with our chosen Hamiltonian.  For example,
the relative size of gaps between single particle states can be used to determine where
shell closures are predicted. 

We select the CD-Bonn NN interaction \cite{Mach1,Mach2} and include 
the Coulomb interaction
between the protons. For $^4He$ we employ the 1996 CD-Bonn 
\cite{Mach1} while for $^{16}O$ we employ the 2000 CD-Bonn \cite{Mach2}.  Where 
comparisons exist, the differences in these interactions are minor and 
are not expected to influence the results of our investigations.

Our selection of model space sizes and harmonic oscillator basis parameter are as follows:
for both nuclei we conduct the SHF evaluations in a model space of 6 major shells.  In addition, for
$^{16}O$ we also provide results for model spaces of 4 and 5 major shells.  
For the NCSM, we select
$N_{max}$ = 10 ($^4He$) and $N_{max}$ = 6 ($ ^{16}O$).
For  $^4He$ we use $\hbar \Omega=22$ MeV while for  
$^{16}O$ we use $\hbar \Omega=15$ MeV.

By way of explanation of the differences in model spaces sampled by SHF and by NCSM, 
we note that the model spaces are selected for a precise NCSM calculation with the 
evaluated $H_{eff}$.  As a result, the SHF calculation samples a somewhat different 
basis space where all nucleons are allowed to be excited through a set of single particle basis states
depending on the number of oscillator shells included.   Our philosophy is to fix the $H_{eff}$ 
and to use it for
both the NCSM and SHF applications.   In this way, we test how well the SHF results approximate the NCSM
results, where the  NCSM results are expected to converge to the exact answers as the model space size
increases.    Fig. 1 displays the two-particle model spaces employed in the SHF and NCSM for $^4He$
and
$ ^{16}O$.   In areas where two-body effective Hamiltonian matrix elements are required for 
SHF but are absent in the NCSM, we simply use the relative kinetic energy matrix elements.

We solve for the properties of our selected nuclei using the SHF code underlying 
the results of Refs. \cite {c5, c6, c7, c8} and the
$m-$scheme Many-Fermion Dynamics code (MFD) \cite{MFD} which was
developed for NCSM calculations. 

In order to gauge the overall effectiveness of the SHF method, we also evaluate 
the second order perturbative corrections to many observables presented here.  
We find the perturbative corrections significantly improve the agreement between 
SHF and NCSM for applications as may be expected for closed shell systems. We label our
perturbatively corrected results with "SHF(2)".

Let us specify the occupied SHF orbitals by capital Roman letters and the unoccupied 
SHF orbitals by lower case Roman letters.  We allow for full charge dependence so 
that neutron and proton orbitals are separately indicated and we employ $\epsilon$ 
to signify the self-consistent SHF single-particle energies.  Then, for example, 
we evaluate the second order correction to the SHF binding energy:

\begin{eqnarray}
\Delta E_{SHF} &=& - \sum _{J, i \le j, A \le B} (2J+1) \frac {\left [ \langle A_nB_n,J | H_{eff} | i_nj_n,J \rangle
\right ]^2}{\epsilon _{i_{n}} + \epsilon _{j_{n}} - \epsilon _{A_{n}} - \epsilon _{B_{n}}}\\
&& - \sum _{J, i,j, A,B} (2J+1) \frac {\left [ \langle A_nB_p,J | H_{eff} | i_nj_p,J \rangle \right
]^2}{\epsilon _{i_{n}} + \epsilon _{j_{p}} - \epsilon _{A_{n}} - \epsilon _{B_{p}}}\nonumber \\
&& - \sum _{J, i \le j, A \le B} (2J+1) \frac {\left [ \langle A_pB_p,J | H_{eff} | i_pj_p,J \rangle
\right ]^2}{\epsilon _{i_{p}} + \epsilon _{j_{p}} - \epsilon _{A_{p}} - \epsilon _{B_{p}}}\nonumber
\end{eqnarray}

\noindent where we signify reduced jj-coupled two-body matrix elements of the effective Hamiltonian in the
SHF basis by $ \langle a b,J | H_{eff} | c d, J \rangle$.    In this shorthand notation, a, b, c, and d represent
either occupied or unoccupied neutron or proton states.  Our two-body states are normalized and
antisymmetrized.

Furthermore, we evaluate the second order correction to the occupation probabilities.  We present the
change in the occupation probability $\Delta N_{A_{n}}$ of the occupied SHF neutron orbital $A_n$ as an
example from which other cases ($\Delta N_{A_{p}}$, $\Delta N_{i_{n}}$, $\Delta N_{i_{p}}$) can easily be
determined by appropriate modifications.

\begin{eqnarray}
\Delta N_{A_{n}}&=& \sum _{J, B, i \le j} (2J+1) \frac {\left [ \langle A_nB_n,J | H_{eff} | i_nj_n,J \rangle
\right ]^2}{(\epsilon _{i_{n}} + \epsilon _{j_{n}} - \epsilon _{A_{n}} - \epsilon _{B_{n}})^2}\\
&& + \sum _{J, B, i, j} (2J+1) \frac {\left [ \langle A_nB_p,J | H_{eff} | i_nj_p,J \rangle \right
]^2}{(\epsilon _{i_{n}} + \epsilon _{j_{p}} - \epsilon _{A_{n}} - \epsilon _{B_{p}})^2} \nonumber
\end{eqnarray}

In every case, we verify by direct evaluation, that the number of neutrons and the number of protons is
separately conserved through the second order calculations.

The second order corrections to the SHF one-body density are easily evaluated from the corrections to the
occupation probabilities.  These corrections, in turn lead to a second order correction to the rms radius.

We also introduce a standard correction to the SHF one-body density \cite {Negele70}
to adjust the rms radius (RMS) for the spurious center-of-mass motion.  This correction is defined as
$RMS = [(RMS_{SHF})^2 - b^2/A]^{1/2}$.  For $A=4$, and $16$ we use $b=1.374~fm$
and $1.663~fm$ respectively.  All theoretical results for rms radii quoted here are for pointlike 
nucleons - i.e we do not adjust for an electromagnetic radius of the nucleons.  The sequence of corrections
presented below begins with the second order perturbative correction to the rms radius.  We then 
apply the center-of-mass motion correction (quoted separately in the tables as $\Delta Spur(cm)$) to our
SHF(2) result. The rms radius resulting from these corrections is also quoted in the tables below as the "Total".

In the following sections, we investigate several observables as well as
properties of the wavefunctions.  For observables, we evaluate the binding energy of the $GS$ 
as well as its
$r_{rms}$, one-body density distributions, single-particle energies  and occupation probabilities.

\section{Application to $^4He$}

In this section we apply the methods outlined in Sec. 2 to 
evaluate the properties of $^4He$ in the SHF and NCSM approaches. 

For the NCSM, we use a complete $N_{max} \hbar \Omega$ model space with $N_{max}=10$ for the 
positive-parity states. This means that a total of eleven major harmonic-oscillator 
shells are involved. The two-nucleon model space shown in Fig. 1, is then defined 
by $N_{max}$, such that the restriction of the harmonic-oscillator single particle
states is given by $N_1 = 2n_1+l_1 \leq N_{max}$, $N_2 =2n_2+l_2 \leq N_{max}$ and 
$(N_1 + N_2) \leq N_{max}$. Thus, the maximum excitation of two nucleons simultaneously 
is through the sixth shell.  

For the SHF we have a cutoff in basis states for each orbital - i.e. for each $(lj)$ pair. 
Clearly, we would have to make some arbitrary choices if the SHF
model space covers areas exceeding  the NCSM space.  We do make some attempts at this below
where we examine  the sensitivity to the choice of the SHF basis in more detail with the applications to 
$^{16}O$.
However, for $^4He$ we have selected the largest SHF basis included entirely within the 
$N_{max}=10$ space of the NCSM.  Thus, we select the 6-shell SHF basis shown in Fig. 1 
that includes  three S-states, three P-states, two D-states, etc. 

On the other hand, there is a range of two-body matrix elements of $H_{eff}$ which participate in the NCSM
calculations but not in the SHF calculations.  To a certain degree, the present work tests the importance of
those matrix elements for the $GS$ properties of $^4He$.  

In Table I we present the experimental $GS$ properties \cite{he4exp, ndata} 
along with the corresponding theoretical results  from SHF and NCSM. 

First, we note that the NCSM ground state energy is near the typical underbound result obtained with realistic
interactions.  In fact, the converged binding energy with the CD-Bonn
NN potential is -26.30(15) MeV obtained in calculations that employ basis 
spaces up to $N_{max}=18$ \cite{Petr00}.  At the same time, the SHF result appears to be rather far from
NCSM with 13.76 MeV less binding.  Most of this difference is recovered with  the second order corrections to
SHF, that is the results of SHF(2), leaving a net 2.9 MeV difference.  While the SHF(2) is considerably
closer to the NCSM, the second order correction may raise concern over the overall rate of convergence
of the perturbative corrections to SHF for  $^4He$.  We note, however, that the second order correction is only
15.3\% of the total SHF interaction energy [-70.946 MeV].  Hence, it appears reasonable to expect most of the
remaining difference between SHF(2) and NCSM will be obtained in third order.

For the rms radii, the NCSM and SHF are rather close to each other and to
experiment where it is available.  The agreement between SHF(2) and NCSM is especially satisfying once the
SHF(2) rms radius is corrected for spurious c.m. motion as described above.

In Table 2 we present occupation probabilities for selected orbitals.  It is important to note that we present
some results in the SHF basis and some in the HO basis.  In particular, the first two columns present the
probabilities that the neutron and proton orbitals are described by pure HO orbitals.  The last two columns
present the HO single particle state occupation probability from the NCSM wavefunction.  The intermediate two
columns present the second order correction to the ground state occupation probabilities {\it in the SHF basis}.

For $^4He$, we observe rather good overlap of the SHF
ground state with the lowest HO configuration.  In addition, the NCSM indicates a rather pure HO lowest
configuration description of the ground state.  Hence there is overall close agreement.  This agreement is
retained since the SHF(2) corrections for the occupied $S_{1/2}$ orbital appear to be rather small.  The
unoccupied SHF orbitals indicate a strong degree of mixing but they do not directly contribute to the ground
state SHF energy so this mixing is less relevant.  

Overall, one feature is rather noticeable - the SHF wavefunction is less "correlated" than the NCSM ground state
wavefunction.  In the SHF basis, SHF(2) is encouraging with its trend indicating correlation mixtures
approaching those of NCSM for the 0S orbital.  At first glance, this comparison may
appear a little dangerous as we are comparing results in a SHF basis with those in a HO basis. However, since
the SHF occupied orbital is largely dominated by a single HO orbit, the differences from the SHF occupation
probabilities  expressed in the HO basis would be negligible.  This is the case, in general, for both light nuclei
treated in the present work.

In Table 3 we present the SHF single particle energies for $^4He$. We note again that the single 
particle energies cannot be compared directly with the experimental separation energies
without considering the expected large rearrangement effects.  

Our spectrum of single particle energies is shifted about 20 MeV from the results obtained with
phenomenological Hamiltonians that include a one-body part.  This shift has been addressed \cite {Jaqua92}
in some detail and is related to the role of the c.m. motion.  Our single particle energies contain a contribution
of approximately ${ \langle T_{rel}/A \rangle}$ where the expectation value is with respect to the
self-consistent single particle wavefunction.  

Recalling the factor of one-half discussed in the preceeding section,
we may interpret the results of Table 3 as predicting a $0P_{3/2}-0S_{1/2}$ particle-hole excited state of
$^4He$, neglecting rearrangement effects, at about 34.6/2 = 17.3 MeV. This is low compared to the 
lowest negative parity excited state involving this configuration at 21.84 MeV of excitation.

We present the radial one-body density distributions for $^4He$ in Figure 2.  One is struck by the apparent
large differences between the mean field, either SHF or SHF(2), and the NCSM distributions.   Overall, the
distributions presented are similar in shape but appear scaled by an amount indicated by their rms radii (see
Table 1).  We shall see below that such a simple scaling does not appear in $^{16}O$.

One major difference between our mean field radial distributions compared with NCSM is
due to a spurious center-of-mass smearing effect present in our mean field results.  We
anticipate that, as we proceed to heavier systems, one of our major goals, this spurious 
effect will be less significant.

Some differences between SHF and NCSM results are due to the different model spaces using the $10
\hbar \Omega$ NCSM effective Hamiltonian derived for $^4He$ as depicted in Fig. 1.  Below, we will investigate
the significance of different model spaces using the $^{16}O$ case with a $6 \hbar \Omega$ effective
Hamiltonian derived for the NCSM model space.

\section{Application to $^{16}O$}

For $^{16}O$, we conduct the NCSM investigations in a 6 $\hbar \Omega$
model space, the largest that is currently feasible for this nucleus. In the m-scheme, for $M=0$
configurations, the 6$\hbar\Omega$ (8$\hbar\Omega$) basis dimensionality is 26,483,625 (996,878,170) for
this nucleus.  We conduct the SHF calculations in a series of three model spaces (4-shells, 5-shells and 6-shells)
that cover a range of situations both smaller and larger in certain aspects than the NCSM model space.  These
selections are compared in Figure 1.

In the 6-shell space, the SHF is missing certain matrix elements due to the
limitations of the NCSM model space.  This corresponds to the region where the SHF model space contains
two-nucleon excitations beyond the NCSM model space. When this occurs in our SHF calculations, the $V_{eff}$
matrix elements vanish while we do retain the unrenormalized relative kinetic energy matrix
elements.

In Table 4 we present the
experimental and theoretical ground state properties for $^{16}O$.  We examine the dependence of 
the SHF results on the number of shells included in such a way as to bracket
the division in model space accomplished in the corresponding NCSM results as shown in Fig.1.

The SHF ground state energy is between 7
and 25 MeV above the NCSM result.   When we include the SHF(2) corrections, the differences are altered to a
range of 6 to 32 MeV.  The larger these corrections, the more significant they are as indicators of possible
difficulties with a convergent perturbation theory based on SHF for this nucleus.  However, when viewed on
the scale of the total interaction energy, these concerns are reduced. The second order correction is
[6.5\%, 10.1\%, 6.9\%] of the total SHF interaction energy [-481.01, -492.37, -554.61] MeV in the [4, 5, 6] shell
model spaces respectively.  In all these $^{16}O$ cases, the small percentage change when including second
order corrections is encouraging for our goal of treating heavier systems in SHF(2).

We note from the binding energies in Table 4 that the SHF results are closest to NCSM in the 6 shell case while,
with second order corrections, the 4 shell results are closest.

Let us also address the issue of convergence by comparing the size of our second order corrections in
$^{16}O$ with the perturbative corrections obtained in Ref. \cite {Coraggio03} using a different approach and
featuring a realistic smoothed nucleon-nucleon interaction, $V_{low-k}$ \cite {Bogner}.
They obtain a second order correction that is 17\% of their total SHF interaction energy of -376 MeV and a
third order correction of 8\%.  We note that the average of our second order corrections (7.8\%) is comparable
in percentage to their third order correction.  

Table 5 presents for $^{16}O$ the occupation probabilities for selected orbitals in the 6-shell SHF
calculations and compares them with the $N_{max}=6$ results of the NCSM. We see that SHF shows
greater mixing in the SHF occupied orbits than does NCSM which provides a distinctive situation from that
observed above for $^4He$. 

The second order corrections to the SHF occupation probabilities presented in Table 5 are all
small and consistent with a well-behaved perturbation theory. As a figure of merit, we note that the total
neutron and proton percentage promoted from occupied to unoccupied orbits is about 5\% in the 4-shell case
while increasing slightly in the 5-shell and 6-shell SHF results.  Hence, the concern raised above with
the apparent large second order corrections to the SHF energy are again reduced.

We present the $^{16}O$ single particle energies in Table 6 for the 4, 5 and 6 shell SHF results.  We again
cite our warning about direct comparison between these single particle energies and experimental states in
odd mass neigboring nuclei.  

First, we note that the Hartree Fock energy difference
for the neutron orbits, $0D_{5/2} - 0P_{1/2}$, is
$( 23.6, 22.4,  24.7)/2 = ( 11.8, 11.2, 12.4) $ MeV in the
4, 5 and 6 shell results respectively.  These values
compare favorably with the experimental
neutron $0D_{5/2} - 0P_{1/2}$ splitting of $11.52 MeV$
obtained from the binding energy differences of
$^{17}O$ and $^{15}O$. Similar results are obtained
when comparing the proton SHF single particle energies
with experimental binding energies of neighboring
odd nuclei after accounting for Coulomb corrections.
The proton $0D_{5/2} - 0P_{1/2}$ splittings are
$( 23.3, 22.4,  24.5)/2 = ( 11.7, 11.2, 12.3) $ MeV
in the 4, 5 and 6 shell results respectively. The
relevant experimental splitting is $11.53 MeV$.
Thus, the size of the gap between the occupied and
unoccupied SHF states that we find in $^{16}O$
is in accord with the known doubly-magic
character of this nucleus.  We also note that
Ref. \cite {Coraggio03} obtains the corresponding (full)
single particle energy splitting of 15.6 MeV.

In a similar vein, and with similar caution, we may examine our spin-orbit splittings.  For example, 
the 4, 5 and 6-shell neutron
$0P_{3/2}-0P_{1/2}$ splittings of Table 6 are $(10.0,  10.9, 13.9)/2 = (5, 5.5, 7) $ MeV respectively, which 
are in approximate agreement with the experimental splitting in $^{15}O$ of $6.2$ MeV.  
We note that Ref. \cite {Coraggio03} obtains a spin-orbit splitting of $7.6$ MeV for these P-states. 

A corresponding comparison of the neutron $0D_{5/2}-0D_{3/2}$ splittings yields $(9.6, 10.4, 11.5)/2 = (4.8,
5.2, 5.8)$ MeV in comparison with the experimental splitting of $5.1$ MeV and the result of $5.9$ MeV in Ref. 
\cite {Coraggio03}.

In Figure 3 we present the radial one-body density distributions for $^{16}O$ obtained in the NCSM and SHF
calculations.  Here we note significant differences between SHF and NCSM, especially in the central region.  It is
worth commenting that our SHF results are quite consistent with
long-established results of Brueckner Hartree-Fock \cite {Negele70} and Coupled Cluster \cite {KLZ78}.  In fact
our NCSM results are somewhat closer to the traditional results from density dependent Hartree Fock (DDHF)
either with phenomenological interactions \cite {VB72} or with higher order Brueckner approaches such as
Renormalized Brueckner Hartree-Fock (RBHF) \cite {Davies71}.  Hence, the more surprising result is the NCSM
smooth gaussian-like shape (solid line).  This implies that simple scaling cannot reduce the differences
between SHF and NCSM in the case of $^{16}O$.

\section{Conclusions and outlook}

We have compared results obtained with exact diagonalization in large 
multi-shell model spaces ($ab~ initio$ no-core shell model) with the approximate 
results from spherical Hartree-Fock using realistic effective two-body Hamiltonians.  
Significant differences are obtained and second order corrections to SHF 
bring the SHF into reasonable agreement with NCSM for $^4He$ and $^{16}O$ in SHF model spaces "enclosed"
by the NCSM model space.  By "enclosed" we refer to the sketch of model spaces in Figure 1 where the
meaning is clear from the labelled model spaces.

One recent effort \cite {SSR00}, with which we can compare our results, shows
that higher order corrections to Hartree-Fock using phenomenological
interactions are significantly smaller than those we obtain here.  It is 
reasonable, in our view, that the 
rates of convergence of higher order corrections to SHF are different between 
realistic effective Hamiltonian approaches and phenomenological interactions.

These phenomenological interactions have been adjusted within Hartree-Fock to 
provide a good description of many experimental observables using the mean 
field approximation.  For the NCSM, we now understand that residual differences 
between theory and experiment in light nuclei are due to contributions from effective 
and real three-body forces.  How possible differences between NCSM theory and 
experiment will be resolved in heavier systems will require further investigation.

Another recent effort \cite {Coraggio03}, with which we can also compare our results, shows
somewhat larger higher order corrections to Hartree-Fock using a different realistic
effective Hamiltonian. The resulting mean field excitation spectra of $^{16}O$ are rather 
similar considering the differences in our approaches.  With the caveat that rearrangement
corrections are not included, both approaches give spin-orbit splittings in rough accord with
experiment.  Our mean field rms radii are somewhat smaller than those of Ref. \cite {Coraggio03} 
and smaller than experiment but our mean field rms radii approximately agree with the NCSM results.  
This raises additional questions regarding the different mean field treatments of the c.m. motion.

Additional questions worth examining in the future include making a similar comparison between SHF and
NCSM with effective three-body Hamiltonians including true three-body forces.  It is anticipated that such
additional study will be especially worthwhile if the expected improved agreement between theory and
experiment with realistic effective Hamiltonians is achieved.

We also conclude that investigations of heavier closed shell nuclei with SHF(2) are now warranted where the
NCSM results are not obtainable in the near future.

\section{Acknowledgements}

M.A. Hasan and J.P. Vary acknowledge 
partial support from  NSF Contract INT00-80491.
J.P. Vary acknowledges support from USDOE Grant No. DE-FG02-87ER-40371.  
This work was performed in part under the auspices of
the U. S. Department of Energy by the University of California,
Lawrence Livermore National Laboratory under contract
No. W-7405-Eng-48. P.Navratil acknowledges support from LDRD
contract 00-ERD-028.

\pagebreak

\pagebreak

{\large{\bf {Figure Captions\\}}}

{\bf {Fig. 1:}} (Color online) Depictions of the various $P_2$ space projectors
defining the model spaces employed in the SHF and NCSM calculations.\\

{\bf {Fig. 2:}} (Color online) One-body radial density distributions obtained
 in the SHF and NCSM calculations for $^4He$.\\

{\bf {Fig. 3:}} (Color online) One-body radial density distributions obtained
 in the SHF and NCSM calculations for $^{16}O$.\\

\pagebreak

\begin{table}  
\caption{Experimental and calculated observables for
the ground state of  $^4He$
with an $N_{max}=10$ effective Hamiltonian based on the 1996 CD-Bonn \cite {Mach1} and using $\hbar
\Omega=22~MeV$. Experimental and calculated ground state energy (in $MeV$) and rms radii (in fm).
The (negative) correction for spurious center-of-mass motion ("$\Delta$ Spur(cm)")is described in the text. For
the experimental rms radius, we take the measured charge radius and correct for the contribution of the
proton charge rms radius(0.8 fm).}

\begin{tabular}{|c|c|c|c|c|}
\hline                                                   
Observable & Experiment & SHF  &  SHF + $\Delta$ SHF & NCSM\\
                  &                    &$\Delta$ SHF & +$\Delta$ Spur(cm)& \\
                  &                    &$\Delta$ Spur(cm)& &\\
\hline                                                   
$E_{GS}$ & -28.296 & -14.156 & -24.991 & -27.913 \\
              &              &  -10.835 &              &             \\
\hline
$n-rms$ &              &    1.584  &               &  1.411  \\ 
              &              &               &               &            \\
$p-rms$ & 1.450    &    1.590  &                &  1.416 \\
              &              &               &                &           \\
$rms$    &              &    1.587   &  1.560     &  1.413 \\ 
              &              &    0.118   &                &           \\
              &              &   -0.145  &                &            \\
\hline
\end{tabular}  
\end{table}

\begin{table}  
\caption{Selection of occupation probabilities for
the ground state of  $^4He$.
The columns "SHF" and "NCSM" label the probabilities in the HO basis.
Specifically, in the case of the SHF unoccupied orbits, we quote their expansion
probabilities in the HO basis.
The $\Delta SHF$ columns present the second order perturbative corrections in
the SHF basis.}

\begin{tabular}{|c|cc|cc|cc|}
\hline                                                   
Orbital & SHF & SHF  & $\Delta$ SHF & $\Delta$ SHF & NCSM & NCSM\\
            &neutron&proton&neutron&proton&neutron&proton\\
\hline                                                   
$0S_{1/2}$& 0.992 	 &  0.991  &  -0.050  &  -0.050  & 0.941 &  0.940  \\
$1S_{1/2}$& 0.005    &  0.005  &  0.002  &  0.002  & 0.008 &  0.008  \\
$2S_{1/2}$& 0.004    &  0.005  &  0.000  &  0.000  & 0.003 &  0.003  \\ 
$0P_{3/2}$& 0.794 	&   0.784  &  0.007  & 0.007  &  0.004 &  0.004  \\
$0P_{1/2}$& 0.629 	&   0.621  &  0.014  &  0.015 &  0.016 &  0.016  \\
$0D_{5/2}$& 0.833 	&   0.829  &  0.000  &  0.000 &  0.001 &  0.001  \\
$0D_{3/2}$& 0.764 	&   0.761  &  0.002  & 0.002  &  0.002 &  0.002  \\
\hline
\end{tabular}  
\end{table}

\begin{table}  
\caption{SHF single particle energies for  the ground state of  $^4He$
in a 6-shell model space using $\hbar \Omega=22~MeV$.}

\begin{tabular}{|c|cc|}
\hline                                                   
Orbital &neutron&proton\\
\hline                                                   
$0S_{1/2}$& -7.546 	 &  -6.610    \\
$0P_{3/2}$& 27.156 	&   28.218   \\
$0P_{1/2}$& 30.409	 &   31.367   \\
$0D_{5/2}$& 40.877   &   41.855  \\
$1S_{1/2}$& 35.955   &   36.936   \\
$0D_{3/2}$& 43.003   &   43.947   \\ 
\hline
\end{tabular}  
\end{table}

\begin{table}  
\caption{Experimental and calculated observables for
the ground state of  $^{16}O$
with an $N_{max}=6$ effective Hamiltonian based on the 2000 CD-Bonn \cite {Mach2} and using $\hbar
\Omega=15~MeV$. Experimental and calculated ground state energy (in $MeV$) and rms radii (in fm).
The (negative) correction for spurious center-of-mass motion ("$\Delta$ Spur(cm)") is described in the text. For
the experimental rms radius, we take the measured charge radius and correct for the contribution of the
proton charge rms radius(0.8 fm). SHF results are presented for 4-shell, 5-shell and 6-shell model spaces. }

\begin{tabular}{|c|c|c|c|c|}
\hline                                                   
Observable &  4-shell SHF     & 5-shell SHF     & 6-shell SHF     & NCSM\\
$[Experiment]$ & $\Delta$ SHF &$\Delta$ SHF    &$\Delta$ SHF   &\\
                        & $\Delta$ Spur(cm)& $\Delta$ Spur(cm)& $\Delta$ Spur(cm) \\
                  &  Total               & Total               & Total              &\\
\hline                                                   
$E_{GS}$ &  -107.46 & -109.83 & -126.00 & -132.87\\
$[-127.62]$&    -31.46 &   -49.88 &   -38.21 &             \\
              &  -138.92 & -159.71 & -164.21 &             \\
\hline
$n-rms$ &      2.093 &     2.071 &    1.954    &   2.209 \\ 
              &                &              &                 &             \\
$p-rms$ &       2.101 &    2.080 &    1.968    &   2.223 \\
 $[2.58]$    &                 &              &               &             \\
$rms$    &       2.097 &   2.076  &    1.961     &   2.216 \\ 
              &       0.072 &   0.112  &    0.117     &              \\
              &     -0.040 &  -0.040  &   -0.042    &             \\
              &       2.129 &  2.148   &     2.036    &             \\
\hline
\end{tabular}  
\end{table}

\begin{table}  
\caption{Selection of occupation probabilities 
for the ground state of  $^{16}O$
with an $N_{max}=6$ effective Hamiltonian using $\hbar \Omega=15~MeV$.
The SHF calculations were performed in the 6-shell space.
The column "SHF" and "NCSM" labels the
probabilities in the HO basis. Specifically, in the case of the SHF unoccupied orbits, 
we quote their expansion probabilities in the HO basis.
The $\Delta SHF$ labels the second order perturbative corrections in
the SHF basis.}

\begin{tabular}{|c|cc|cc|cc|}
\hline                                                   
Orbital & SHF & SHF  & $\Delta$ SHF & $\Delta$ SHF & NCSM & NCSM\\
            &neutron&proton&neutron&proton&neutron&proton\\
\hline                                                   
$0S_{1/2}$&  0.887	 &  0.892   &  -0.020  &  -0.019  &  0.959 & 0.961 \\
$1S_{1/2}$&  0.102	 &  0.098	 &  0.015  &  0.014  &  0.033 & 0.031 \\
$2S_{1/2}$&  0.012	 &  0.011   &  0.000  &  0.000  &  0.003 & 0.003 \\
$0P_{3/2}$&  0.843	 &  0.851 	&  -0.041  &  -0.040  &  0.935 & 0.939 \\
$1P_{3/2}$&  0.131	 &  0.125 	&  0.006  &  0.005  &  0.032 & 0.029 \\
$2P_{3/2}$&  0.025	 &  0.024 	&  0.002  &  0.002  &  0.004 & 0.004 \\
$0P_{1/2}$&  0.886	 &  0.895	 &  -0.079  &  -0.079  &  0.938 & 0.941 \\
$1P_{1/2}$&  0.087   &  0.079   &  0.004  &  0.004  &  0.020 & 0.018 \\
$2P_{1/2}$&  0.027 	&  0.026	 &  0.000  &  0.000  &  0.004 & 0.004 \\
$0D_{5/2}$&  0.955	 &  0.965	 &   0.013  &  0.013  &  0.013 & 0.013 \\
$0D_{3/2}$&  0.978	 &  0.964	 &   0.017  &  0.019  &  0.015 & 0.015 \\
\hline
\end{tabular}  
\end{table}

\begin{table}  
\caption{SHF single particle energies for  the ground state of  $^{16}O$
in a 4-shell, 5-shell and 6-shell model spaces using $\hbar \Omega=15~MeV$.}

\begin{tabular}{|c|cc|cc|cc|}
\hline                                                   
Orbital & 4-shell & 4-shell & 5-shell & 5-shell & 6-shell & 6-shell \\
            &neutron&proton  & neutron& proton & neutron& proton  \\
\hline                                                   
$0S_{1/2}$& -41.877 	 &  -37.402   & -44.289    & -39.714  & -49.101 & -44.312 \\
$0P_{3/2}$& -10.148  	&    -5.893   & -10.085    &   -5.778  & -12.334 &  -7.743 \\
$0P_{1/2}$&   -0.129 	 &     4.031    &    0.852    &    5.042  &     1.575 &   5.997 \\
$0D_{5/2}$&   23.437   &    27.335    &  23.277   &   27.450  &  26.261 &  30.537 \\
$1S_{1/2}$&   24.840   &    28.614    &  25.037   &   28.842  &  28.255 &  32.075 \\
$0D_{3/2}$&   33.080    &   36.920    &  33.650   &   37.295  &  37.803 &  40.991 \\ 
\hline
\end{tabular}  
\end{table}

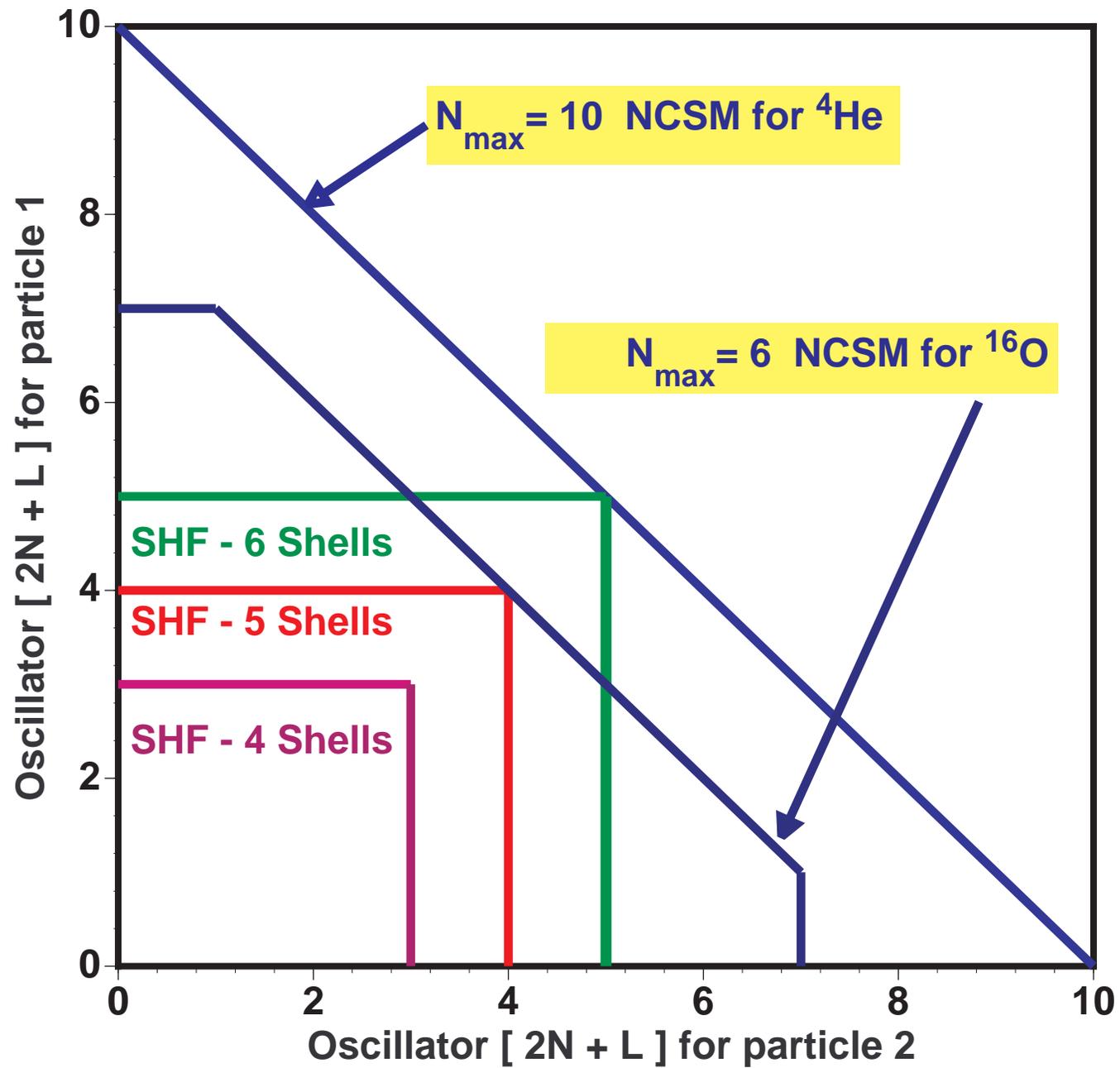
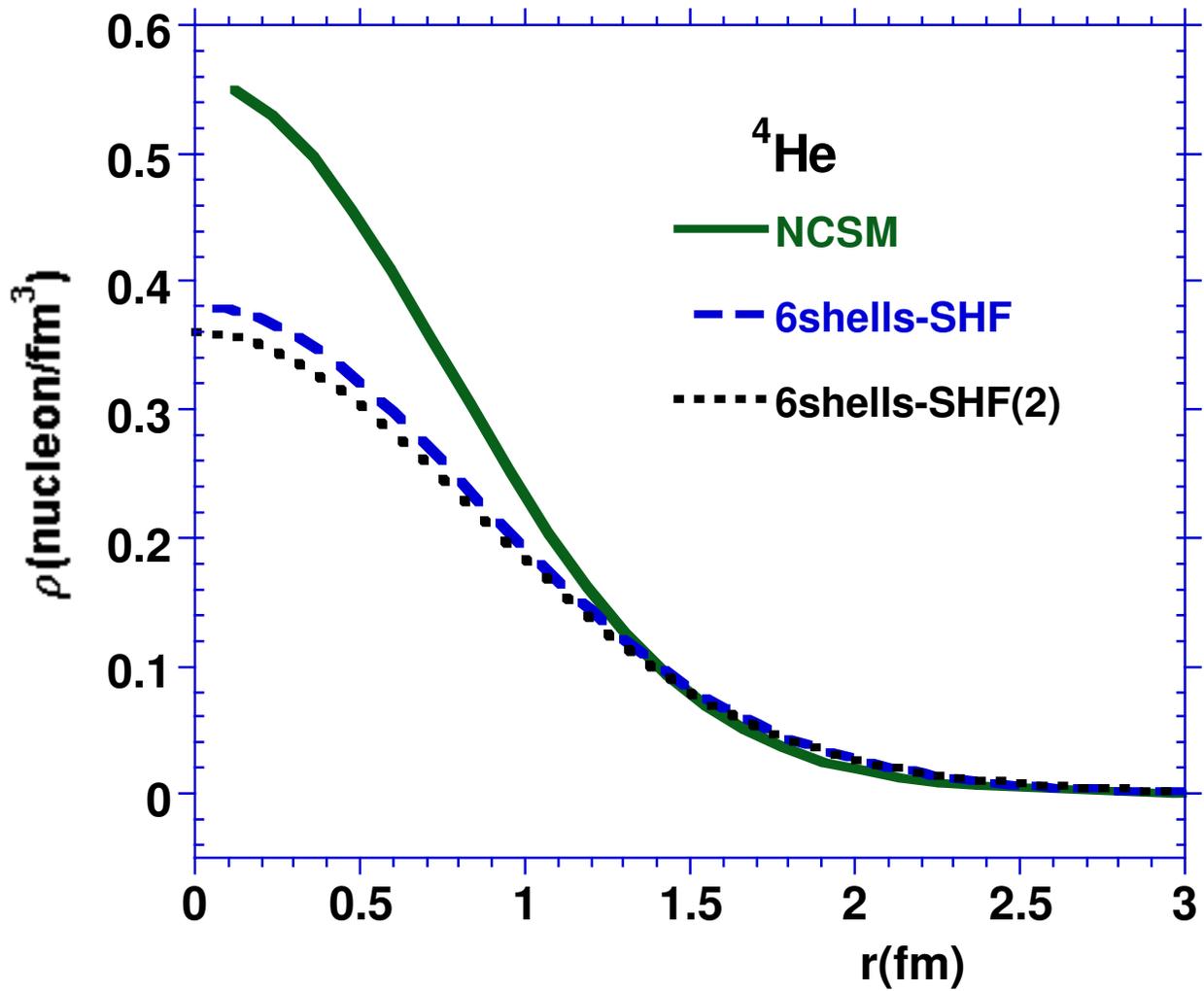
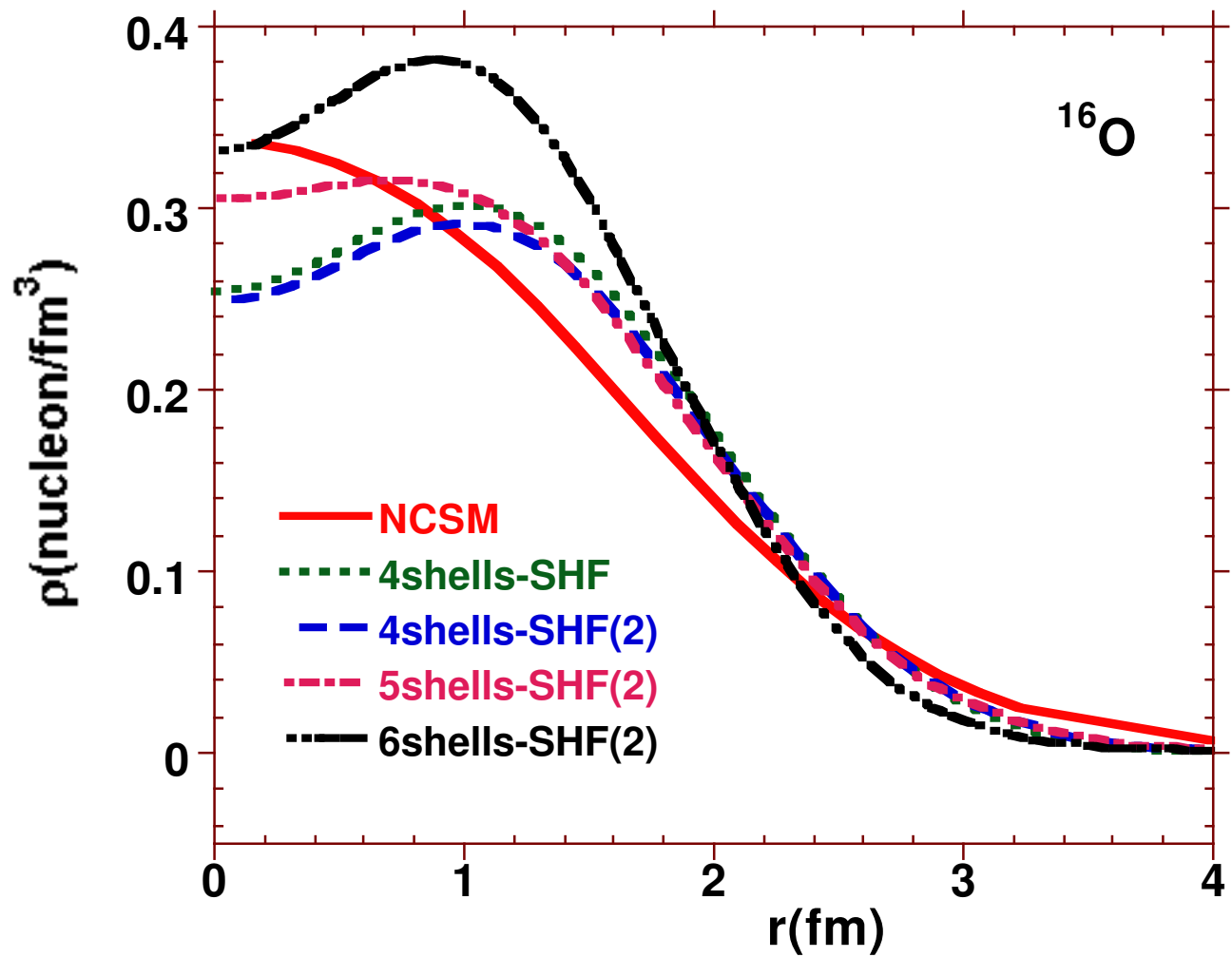
\end{document}